\documentclass[aps,showpacs,amssymb,amsfonts,superscriptaddress,twocolumn,prl]{revtex4-1}  
\usepackage[english,ngerman]{babel}
\usepackage{graphicx}  
\usepackage{dcolumn}   
\usepackage{bm}        
\usepackage{amssymb}   
\usepackage{amsmath}
\usepackage{mathtools}
\usepackage{braket}
\usepackage[utf8]{inputenc}
\usepackage{changes}
\usepackage{times}

\definecolor{myurlcolor}{rgb}{0,0,0.7}
\definecolor{myrefcolor}{rgb}{0.8,0,0}
\usepackage{hyperref}
\hypersetup{colorlinks, linkcolor=myrefcolor,
citecolor=myurlcolor, urlcolor=myurlcolor}

\usepackage{amsthm}

\newcommand{\ignore}[1]{}

 \definecolor{jordi}{rgb}{0.1,0.1,0.5}
  \definecolor{remik}{rgb}{0.5,0.1,0.5}

\def\ketbra#1#2{{\vert#1\rangle\!\langle#2\vert}}

\begin{document}
\selectlanguage{english}

\title{Device-independent witnesses of entanglement depth from two-body correlators}

\author{A. Aloy}
\affiliation{ICFO-Institut de Ciencies Fotoniques, The Barcelona Institute of Science and Technology, 08860 Castelldefels (Barcelona), Spain}

\author{J. Tura} \email{jordi.tura@mpq.mpg.de} 
\affiliation{Max-Planck-Institut f\"ur Quantenoptik, Hans-Kopfermann-Stra{\ss}e 1, 85748 Garching, Germany}

\author{F. Baccari}
\affiliation{ICFO-Institut de Ciencies Fotoniques, The Barcelona Institute of Science and Technology, 08860 Castelldefels (Barcelona), Spain}

\author{A. Ac\'in}
\affiliation{ICFO-Institut de Ciencies Fotoniques, The Barcelona Institute of Science and Technology, 08860 Castelldefels (Barcelona), Spain}
\affiliation{ICREA, Pg. Lluis Companys 23, 08010 Barcelona, Spain}

\author{M. Lewenstein}
\affiliation{ICFO-Institut de Ciencies Fotoniques, The Barcelona Institute of Science and Technology, 08860 Castelldefels (Barcelona), Spain}
\affiliation{ICREA, Pg. Lluis Companys 23, 08010 Barcelona, Spain}

\author{R. Augusiak}
\affiliation{Center for Theoretical Physics, Polish Academy of Sciences, Aleja Lotnik\'ow 32/46, 02-668 Warsaw, Poland}

\date{\today}

\begin{abstract}
We consider the characterization of entanglement depth in a quantum many-body system from the device-independent perspective; that is, we aim at certifying how many particles are genuinely entangled without relying on assumptions on the system itself nor on the measurements performed. We obtain device-independent witnesses of entanglement depth (DIWEDs) using the Bell inequalities introduced in [J. Tura \textit{et al.}, \href{https://doi.org/10.1126/science.1247715}{Science {\bf 344} 1256} (2014)] and compute their $k$-producibility bounds. To this end, we exploit two complementary methods: first, a variational one, yielding a possibly optimal $k$-producible state; second, a certificate of optimality via a semi-definite program, based on a relaxation of the quantum marginal problem. Numerical results suggest a clear pattern on $k$-producible bounds for large system sizes, which we then tackle analytically in the thermodynamic limit. Contrary to existing DIWEDs, the ones we present here can be effectively measured by accessing only collective measurements and second moments thereof. These technical requirements are met in current experiments, which have already been performed in the context of detecting Bell correlations in quantum many-body systems of $5\cdot 10^2 \sim 5 \cdot 10^5$ atoms.
\end{abstract}

\maketitle

\paragraph{Introduction.} Entanglement dwells at the core of quantum physics \cite{HHHH}. Besides being a holistic feature of quantum systems, it is also a resource for nonclassical tasks such as quantum cryptography \cite{Ekert91} or teleportation \cite{Bennett93}, and gives rise to Bell correlations \cite{Bell1964}, which enable stronger, device-independent (DI) quantum information processing \cite{AcinDIQKD, PironioDIQKD}. Entanglement has also proven essential to grasp quantum many-body phenomena \cite{AmicoRMP2008}, and to be key for quantum simulations \cite{KimNature2010, SimonNature2011} and quantum-enhanced metrology \cite{GiovannettiNatPhot2011}, inspiring even the tensor network ansatz \cite{VMCTN}.

From the experimental perspective, spin-squeezed states have been shown to be entangled \cite{Sorensen2001}, and they are typically prepared in large clouds of atoms in different settings such as thermal gas cells \cite{SorensenPRL1999}, atomic ensembles \cite{RobPRL2012} or Bose-Einstein condensates \cite{GrossNature2010, RiedelNature2010}. 
A central objective in experimental quantum physics is thus the generation and certification of entanglement.

Systems with more than two particles can exhibit entanglement in a whole plethora of ways (see, e.g., \cite{GuehnePhysRep2009}), and much effort has been devoted to detecting its strongest form: Genuine multipartite entanglement (GME) \cite{GuehneNJP2010, JungnitschPRL2011, HuberPRAR2011}. However, the technical requirements for GME detection are usually too demanding to be fulfilled in realistic experimental conditions and one is interested, rather, in characterizing the system's so-called entanglement depth \cite{LueckePRL2014} (i.e., the minimal amount of GME particles within the system \cite{kproducibility}).

The usual approaches to entanglement characterization are based on full tomography, in order to measure the reconstructed density matrix against an entanglement witness. However, these approaches suffer from, at least, two caveats. On the one hand, the exponential growth of the Hilbert space description with the particle number renders them impractical in the many-body regime. On the other hand, they require a deep understanding and faithful characterization of the measurements, states and relevant degrees of freedom of the system. This may be problematic because it is well known that wrong conclusions can be drawn if these assumptions fail, even slightly \cite{BancalDIGME, YCLiangEntDepth}. An alternative approach, allowing to circumvent these issues, are device-independent witnesses of entanglement depth (DIWEDs) \cite{BancalDIGME, YCLiangEntDepth, MoroderPRL2013} (see also \cite{Lin2019} for recent developments), which rely only on the observed statistics arising from a Bell-like experiment.

In this work, we present a method to derive DIWEDs from Bell inequalities with two dichotomic observables per party. When these DIWEDs are based on two-body, permutationally invariant Bell inequalities (PIBIs) \cite{SciencePaper, AnnPhys}, their additional structure enables us to reach larger system sizes in comparison with current methodology, even enabling us to draw conclusions in the many-body regime. Furthermore, such PIBIs imply the possiblity of entanglement detection with collective observables such as total spin components and second moments thereof, as it has been done in recent experiments in the context of Bell correlation witnesses \cite{ExperimentBasel, ExperimentKasevich}.

\paragraph{Preliminaries.}
In a multipartite quantum system, entanglement can manifest in different notions and strengths, which is equivalently mapped to quantum states belonging to different separability classes \cite{GuehnePhysRep2009}. To be more precise, let us consider $n$ parties sharing some multipartite state and a partition ${\cal P}^{(k)}$ of $[n]:=\{1, \ldots, n\}$ into $m$ pairwise disjoint, non-empty subsets $\mathcal{A}_i$, each of size at most $k$. We denote such a partition ${\cal P} = \{{\cal A}_1, \ldots, {\cal A}_{m}\}$ and omit the superindex $k$ whenever it is clear from the context. Then, we say that a pure $n$-partite state $\ket{\Psi}$ is \textit{$k$-producible with respect to the partition $\cal P$} if it can be expressed as
\begin{equation}
 \ket{\Psi} = \ket{\phi_1}_{{\cal A}_1}\otimes \cdots \otimes \ket{\phi_{m}}_{{\cal A}_{m}},
 \label{eq:kprodpure}
\end{equation}
where each $\ket{\phi_i}_{\mathcal{A}_i}$ is a pure state corresponding to the group $\mathcal{A}_i$.
We then say that a mixed state $\rho$ is \textit{$k$-producible} if, and only if, it can be expressed as 
\begin{equation}
 \rho = \sum_{\cal P} \lambda_{\cal P} \ketbra{\Psi}{\Psi}_{\cal P};
 \label{eq:kprod}
\end{equation}
i.e., a convex combination ($\sum_{\cal P} \lambda_{\cal P} = 1$, $\lambda_{\cal P}\geq 0$) of projectors onto the states given in \eqref{eq:kprodpure} over different partitions ${\cal P}^{(k)}$. The minimal $k$ for which a given multipartite state $\rho$ admits a
decomposition (\ref{eq:kprod}) is called \textit{entanglement depth}.

A natural tool to certify entanglement in a device-independent way are Bell inequalities. To be more precise, let us consider the simplest multipartite Bell scenario,
in which $n$ parties share a multipartite quantum state $\rho$. On the corresponding subsystem of $\rho$, each party measures one of two dichotomic observables $\mathcal{M}_k^{(i)}$, whose outcomes are labelled $\pm 1$. This scenario is usually called $(n,2,2)$.
Let $M_{k_1, \ldots, k_p}^{(i_1, \ldots, i_p)}$ denote the $p$-body correlation function in which party $i_j$ measures the $k_j$th observable. Then, a multipartite Bell inequality can be written as $I-\beta_C\geq 0$ with $I$ being a linear combinations of such correlations of the generic form
\begin{equation}
 I:= \sum_{p=1}^n \sum_{k_j \in \{0,1\}} \sum_{1 \leq i_1 < \ldots < i_p \leq n} \alpha_{k_1, \ldots, k_p}^{(i_1, \ldots i_p)} M_{k_1, \ldots, k_p}^{(i_1, \ldots, i_p)},
 \label{eq:BI:n22}
\end{equation}
where $\alpha_{k_1, \ldots, k_p}^{(i_1, \ldots i_p)} \in \mathbb{R}$ and $\beta_C$ is the so-called classical bound defined as $\beta_C=\min_{\mathrm{LHV}} I$ with the minimum taken over all local hidden variable (LHV) theories (or, equivalently, by all correlations arising from $1$-producible states).

Violation of Bell inequalities signals entanglement in quantum systems, 
however, does not specify its depth. Our main 
aim here is to go significantly beyond and design 
Bell-like inequalities capable of revealing entanglement depth in multipartite quantum states. Precisely, we
want to obtain inequalities 
$I-\beta_k\geq 0$, where $I$ is given in Eq. (\ref{eq:BI:n22}) while $\beta_k$---the so-called $k$-producible bound---is defined in an analogous fashion as $\beta_C$, but now the optimization is carried over $k$-producible states and dichotomic measurements of, in principle, any local dimension.

It is clear that the computation of $\beta_k$ is
a formidable task, however, in the simplest $(n,2,2)$ scenario considered here, it can be significantly simplified. 
That is, we can follow the reasoning of Ref. \cite{TonerVerstraete}, to see that to find $\beta_k$ it is enough to perform the optimization over $n$-qubit $k$-producible states and local one-qubit traceless observables $\mathcal{M}_k^{(i)}$. We can then assume, without loss of generality, that all the observables are of the form
 ${\cal M}_k^{(i)} = \cos \theta_{i,k} \sigma_x^{(i)} + \sin \theta_{i,k} \sigma_z^{(i)}$, with $\sigma^{(i)}_{x/z}$ being the Pauli matrices acting on site $i$. Denoting 
 by $\boldsymbol{\theta}$ the vector consisting of all $\theta_{i,k}$, we consequently have
\begin{equation}
 \beta_k =  \min_{\boldsymbol \theta, \rho} \mathrm{Tr}[{\cal B}({\boldsymbol \theta}) \rho],
\end{equation}
where $\mathcal{B}(\boldsymbol{\theta})$ is the Bell operator corresponding to a given $I$ and $\rho$ is an $n$-qubit $k$-producible state of the form \eqref{eq:kprod}.
By a convex-roof argument, the above optimization is attained at a pure state of the form  \eqref{eq:kprodpure}, for some partition ${\cal P}^{(k)}$, which means that 
\begin{equation}
 \beta_k =  \min_{\boldsymbol \theta, \ket{\Psi}} \bra{\Psi} {\cal B}(\boldsymbol \theta) \ket{\Psi}.
 \label{eq:optimizationpurestates}
\end{equation}
We then have $I- \beta_k \geq 0$ for any $k$-producible state.

The minimization in \eqref{eq:optimizationpurestates} can in principle be performed exactly, since it can be expressed as a polynomial function satisfying polynomial equality constraints (coming from the normalization of $\ket{\phi_i}$ and $\cos^2 \theta_{i,k} + \sin^2 \theta_{i,k} = 1$). However, the degree of such polynomial grows in general with the number of parties $n$, potentially yielding a vast quantity of local minima, rendering this approach impractical (see \cite{PRATwin} for details).

In order to significantly facilitate our considerations, in particular the computation of $\beta_k$, in this work we study two-body PIBIs of the form
\begin{equation}
 I := \sum_{k \in \{0,1\}} \alpha_k {\cal S}_k + \sum_{k \leq l \in \{0,1\}} \alpha_{kl}{\cal S}_{kl},
 \label{eq:BI}
\end{equation}
where 
\begin{equation}
\displaystyle {\cal S}_k := \sum_{i \in [n]}{M}^{(i)}_k,\qquad\displaystyle {\cal S}_{kl} := \sum_{i\neq j \in [n]}{M}^{(i,j)}_{k,l},
\end{equation}
To determine $\beta_k$, we have envisaged two complementary numerical methods. The first one allows to build a good guess for it whereas the second one's aim is to certify that this guess is the global minimum of \eqref{eq:optimizationpurestates}.

\paragraph{Variational upper bound to $\beta_k$.}
Building upon the so-called see-saw optimization method \cite{PalVertesiSeeSaw, WernerWolfSeeSaw}, we find a local minimum, denoted $\beta_k^U$, that by construction, upper bounds $\beta_k$, i.e., $\beta_k^U \geq \beta_k$. To this end, we fix a partition $\cal P$ and pick random starting measurement settings $\boldsymbol \theta$ and a random $k$-producible state $\ket{\Psi}$. The see-saw method uses the stochastic gradient descent and iterates back and forth between $\boldsymbol \theta$ and $\ket{\Psi}$, keeping the rest of the parameters fixed. Note that the optimization over $k$-producible states cannot be done via a straightforward semi-definite program (SDP) because the tensor product structure makes it nonlinear in the states. However, one can also use a see-saw optimization scheme here by keeping $\boldsymbol \theta$ and $\ket{\phi_{\cal A}}$ fixed for all ${\cal A} \in {\cal P}$ except one, say ${\cal A}'$.

Here the key advantage of two-body PIBIs is clear: the degree of the polynomial resulting from \eqref{eq:optimizationpurestates} is constant. This drastically reduces the amount of local minima of $\braket{\Psi|{\mathcal{B}}(\boldsymbol{\theta})|\Psi}$, which can now be split as
\begin{alignat}{2}
 & \sum_{{\cal A}\in \mathcal{P}} \left( \sum_{k} \alpha_k \underbrace{\braket{\phi_{\cal A}|{\mathcal{B}}_k^{\cal A}|\phi_{\cal A}}}_{\textmd{one-body terms}}+\sum_{k \leq l} \alpha_{kl} \underbrace{\braket{\phi_{\cal A}|{\mathcal{B}}_{kl}^{\cal A}|\phi_{\cal A}}}_{\textmd{same region terms}} \right)\nonumber\\
+&\sum_{{\cal A}\neq {\cal A}'\in \mathcal{P}}\left(\sum_{k\leq l} \alpha_{kl}\underbrace{\braket{\phi_{\cal A}|{\mathcal{B}^{\cal A}_k}|\phi_{\cal A}}\braket{\phi_{{\cal A}'}|{\mathcal{B}^{{\cal A}'}_l}|\phi_{{\cal A}'}}}_{\textmd{crossed region terms}}\right),
\label{eq:Contraction}
\end{alignat}
where $\ket{\phi_{\cal A}}$ has support on the parties forming region ${\cal A} \subseteq [n]$, and we have defined 
\begin{equation}
\displaystyle {\cal B}^{\cal A}_k:= \sum_{i \in {\cal A}} {\cal M}_k^{(i)},\quad\displaystyle {\cal B}^{\cal A}_{kl}:= \sum_{i \in {\cal A}}\sum_{j \in {\cal A}\setminus \{i\}} {\cal M}_k^{(i)}\otimes {\cal M}_l^{(j)}.
\end{equation}
During the state optimization, due to the form of \eqref{eq:Contraction}, one finds $\ket{\phi_{{\cal A}'}}$ as the eigenvector 
corresponding to the minimal eigenvalue of $\tilde {\cal B}_{{\cal A}'}$, where
\begin{eqnarray}
\tilde{\cal B}_{{\cal A}'}&=&\sum_k \alpha_k {\cal B}_k^{{\cal A}'} + \sum_{k \leq l} \alpha_{kl} {\cal B}_{kl}^{{\cal A}'}  \nonumber\\
&&+\sum_{k \leq l} \alpha_{kl} \left( \sum_{{\cal A} \neq {\cal A}'} \braket{{\cal B}_k^{\cal A}}{\cal B}_l^{{\cal A}'}+{\cal B}_k^{\cal A'}\braket{{\cal B}_l^{{\cal A}}}\right),
\end{eqnarray}
with $\braket{{\cal B}_k^{\cal A}} = \braket{\phi_{\cal A}|{\cal B}_k^{\cal A}|\phi_{\cal A}}$. To improve $\boldsymbol \theta$, one can also use the see-saw optimization by fixing the value of all measurement settings except for one party and iterating. By construction, at each iteration one obtains a lower and lower expectation value $\braket{\Psi|{\mathcal{B}}(\boldsymbol \theta)|\Psi}$ until a minimum is found, which we denote $\beta_k^U$.

This method can be applied to any Bell inequality. However, it may lead to poor upper bounds if the problem has many local minima. Fortunately, as shown below, for our choice of a Bell expression $I$ we reach the global minimum.

\paragraph{Certificate of lower bound to $\beta_k$.}

Consider ${\cal B}(\boldsymbol \theta)$ and a partition $\cal P$. Since ${\cal B}(\boldsymbol \theta)$ contains at most two-body operators, given an arbitrary quantum state $\rho$, $\mathrm{Tr}[{\cal B}(\boldsymbol \theta) \rho]$ expresses as
\begin{eqnarray}
 \mathrm{Tr}[{\cal B}(\boldsymbol \theta) \rho]&=&\sum_{{\cal A}\in {\cal P}} \left(\sum_k \alpha_k \mathrm{Tr}[{\cal B}_k^{\cal A} \rho_{\cal A}] + \sum_{k \leq l} \alpha_{kl}\mathrm{Tr}[{\cal B}_{kl}^{\cal A} \rho_{\cal A}]\right)\nonumber\\
 &&+\sum_{{\cal A}\neq {\cal A}' \in {\cal P}}\sum_{k \leq l} \alpha_{kl}\mathrm{Tr}[{\cal B}_k^{\cal A}\otimes {\cal B}_l^{{\cal A}'}\rho_{{\cal A} \cup {\cal A}'}],
\end{eqnarray}
where $\rho_{\cal A}$ is the reduced state of $\rho$ on the subsystems forming $\cal A$. If one restricts $\rho$ to being separable with respect to some partition $\cal P$, as it is the case for the optimal value of $\beta_k$, then one would need to ensure that $\rho_{\cal A \cup \cal A'}$ is separable across the ${\cal A}|{\cal A}'$ cut. It is known, unfortunately, that deciding whether a bipartite quantum state is separable is NP-hard \cite{Gurvits03}, so there is a priori no easy way to enforce this condition. However, to find a lower bound on $\beta_k$, one can relax the separability condition to an efficiently tractable one, such as requiring $\rho_{\cal A \cup \cal A'}$ to satisfy the positivity under partial transposition (PPT) criterion \cite{PeresPPT}, which we denote $\rho_{\cal A \cup \cal A'}^{T_{\cal A}} \succeq 0$.

Therefore, one can find a lower bound $\beta_k^L$ to $\mathrm{Tr}[{\cal B}(\boldsymbol \theta) \rho]$ by solving the following SDP:
\begin{alignat}{3}
 \beta_k^L = \min \ &\mathrm{Tr}[{\cal B}(\boldsymbol \theta) \rho]\nonumber\\
 \mathrm{s. t. }\ & \rho_{\cal A} \succeq 0,\ \rho_{\cal A \cup \cal A'} \succeq 0,\nonumber\\
 &\mathrm{Tr}[\rho_{\cal A}] = \mathrm{Tr}[\rho_{\cal A \cup \cal A'}]=1,\nonumber\\
 &\mathrm{Tr}_{\cal A'}[\rho_{\cal A \cup \cal A'}] = \rho_{\cal A},\nonumber\\
 & \rho_{\cal A \cup \cal A'}^{T_{\cal A}} \succeq 0.
 \label{eq:SdP}
\end{alignat}
Note that a state yielding $\beta_k$ is of the form of \eqref{eq:kprodpure}, which trivially satisfies the SDP conditions \eqref{eq:SdP} as $\rho_{\cal A \cup \cal A'} = \ketbra{\phi_{\cal A}}{\phi_{\cal A}}\otimes \ketbra{\phi_{\cal A'}}{\phi_{\cal A'}}$. However, the feasible set of \eqref{eq:SdP} is clearly larger and contains configurations that do not come from quantum states, as \eqref{eq:SdP} can be seen as a relaxation of the quantum marginal problem. We note that this method is applicable to any Bell inequality built from marginals. 

Hence, by optimizing $\beta_k^L$ for every partition $\cal P$ and measurement parameters $\boldsymbol \theta$, and $\beta_k^U$ over different partitions $\cal P$, one obtains $\beta_k^L \leq \beta_k \leq \beta_k^U$.

\begin{figure}
\begin{center}
\centering\includegraphics[scale=0.4]{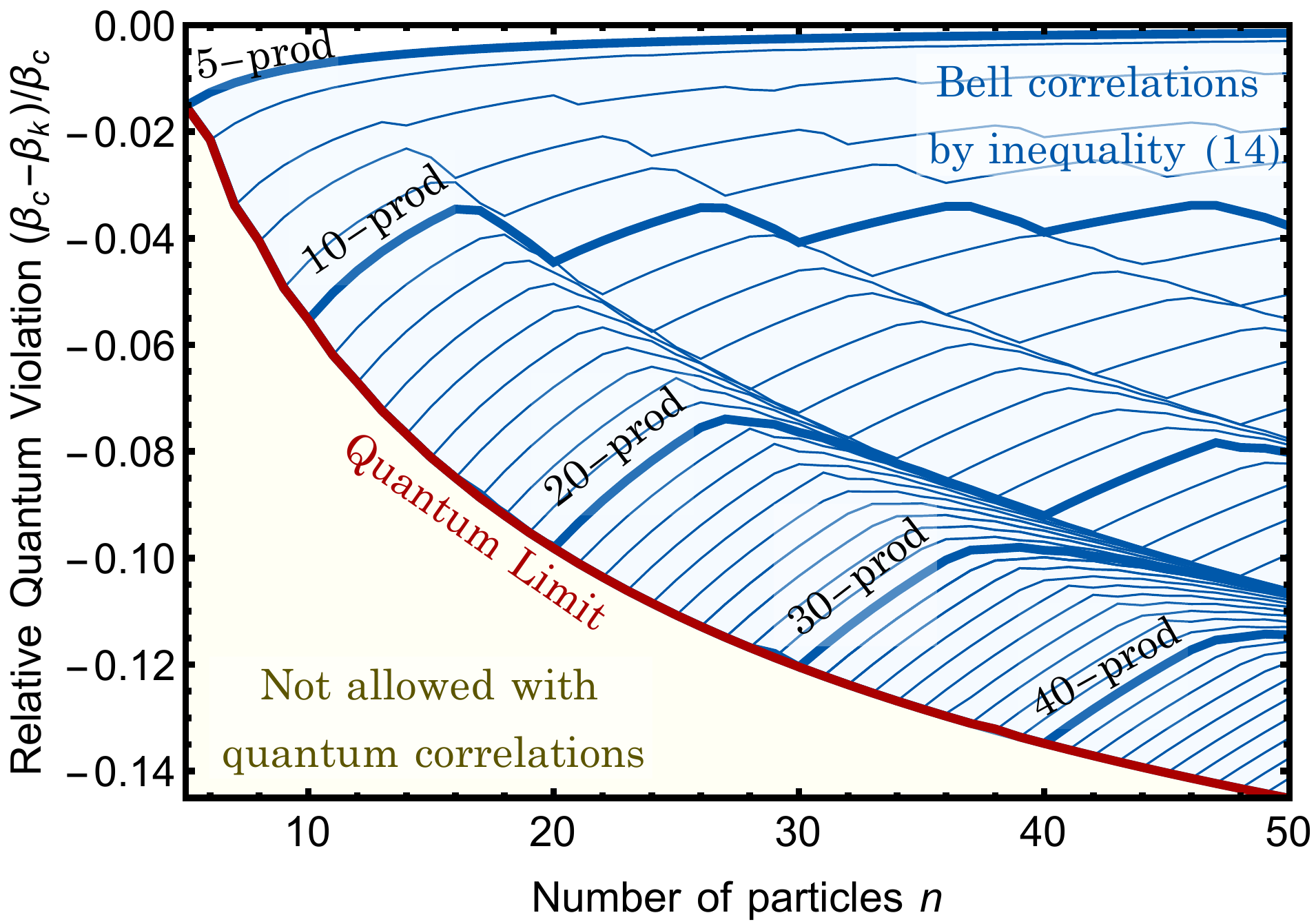}
  \caption{DIWED bounds for the 2-body PIBI \eqref{eq:ineq6} for $n\leq 50$. Each line represents a $k$-producible bound. The wavy-like behavior of the bounds comes from the fact that \eqref{eq:ineq6} has no quantum violation for less than $5$ parties \cite{SciencePaper}. Therefore, the optimal partition $\cal P$ for every pair $(n,k)$ tries to avoid groups of $4$ parties or less. For $n\leq 15$, the optimizations have been performed without assumptions, yielding a gap $\beta_k^U - \beta_k^L$ within numerical accuracy. In the extrapolation for larger $n$, we have assumed the symmetry property within regions ${\cal A}_i$ via \eqref{eq:SchurWeyl} to reduce the number of parameters.}
  \label{fig:waves}
  \end{center}
\end{figure}

\paragraph{Numerical results.} We have seen that the above methods yield values of $\beta_k^U - \beta_k^L$ within numerical accuracy (thus determining $\beta_k$ up to numerical accuracy) for the inequalities introduced in \cite{SciencePaper} (see Fig. \ref{fig:waves}).
Furthermore, there is a strong numerical evidence that for these PIBIs, $\beta_k$ is reached when all parties within each region $\cal A$ pick the same measurement settings (up to local unitary transformations). The two-body structure and the symmetries in the PIBIs greatly reduce the number of local minima in \eqref{eq:optimizationpurestates}. Our methods can explore up to $n=15$ without extra assumptions, limited by the memory requirements of the second method. In \cite{PRATwin} these numerical results are presented in detail.

\paragraph{Extrapolation to the many-body regime.} Numerics suggest that for PIBIs in \cite{SciencePaper} (cf. \eqref{eq:BI}), $\beta_k$ is achieved when the Bell operator becomes invariant with respect to permutations within the regions of the optimal partition $\cal P$. Furthermore, this partition $\cal P$ tends to being the most balanced (i.e., containing as many groups of $k$ parties as possible). As a consequence, one can use Schur-Weyl duality representation theory results \cite{FultonHarris} to split the Hilbert space into invariant subspaces of much smaller dimension, by considering the projector
\begin{equation}
 {\Pi}_{\mathbf{J}}^{\cal P}:= \bigotimes_{\cal A \in {\cal P}} \Pi_{J_{\cal A}}^{\cal A},
 \label{eq:SchurWeyl}
\end{equation}
where $\Pi_{J_{\cal A}}^{\cal A}$ projects the Hilbert space corresponding to $\cal A$ onto the $J_{\cal A}$-th spin length \cite{JordiThesis, MoroderNJP2012}. This is a great simplification, because now it allows us to compute large entanglement depths: recall that $\Pi_{J_{\cal A}}^{\cal A}$ projects the $2^{|\cal A|}$-dimensional subspace onto a $(2 J_{\cal A} +1)$-dimensional subspace, where $J_{\cal A} \leq |{\cal A}|/2$. Interestingly, we also observe that the considered Bell inequalities are always saturated for the maximal spin subspace; i.e., when $J_{\cal A}=|{\cal A}|/2$ for all $\cal A \in \cal P$.

\paragraph{Example.} Let us illustrate our method with an exemplary PIBI (cf. \eqref{eq:BI})
constructed in Ref. \cite{SciencePaper} given by
\begin{equation}
 I = -2 {\cal S}_0 + \frac{1}{2}{\cal S}_{00} - {\cal S}_{01} + \frac{1}{2}{\cal S}_{11}.
 \label{eq:ineq6}
\end{equation}
Fig. \ref{fig:waves} presents the $\beta_k$ to construct the DIWED $I - \beta_k\geq 0$. We have also studied bounds from generalization of CHSH \cite{CHSH} and PIBIs detecting Dicke states \cite{AnnPhys}, which are presented in \cite{PRATwin}.

\begin{figure}
\begin{center}
  \includegraphics[scale=0.45]{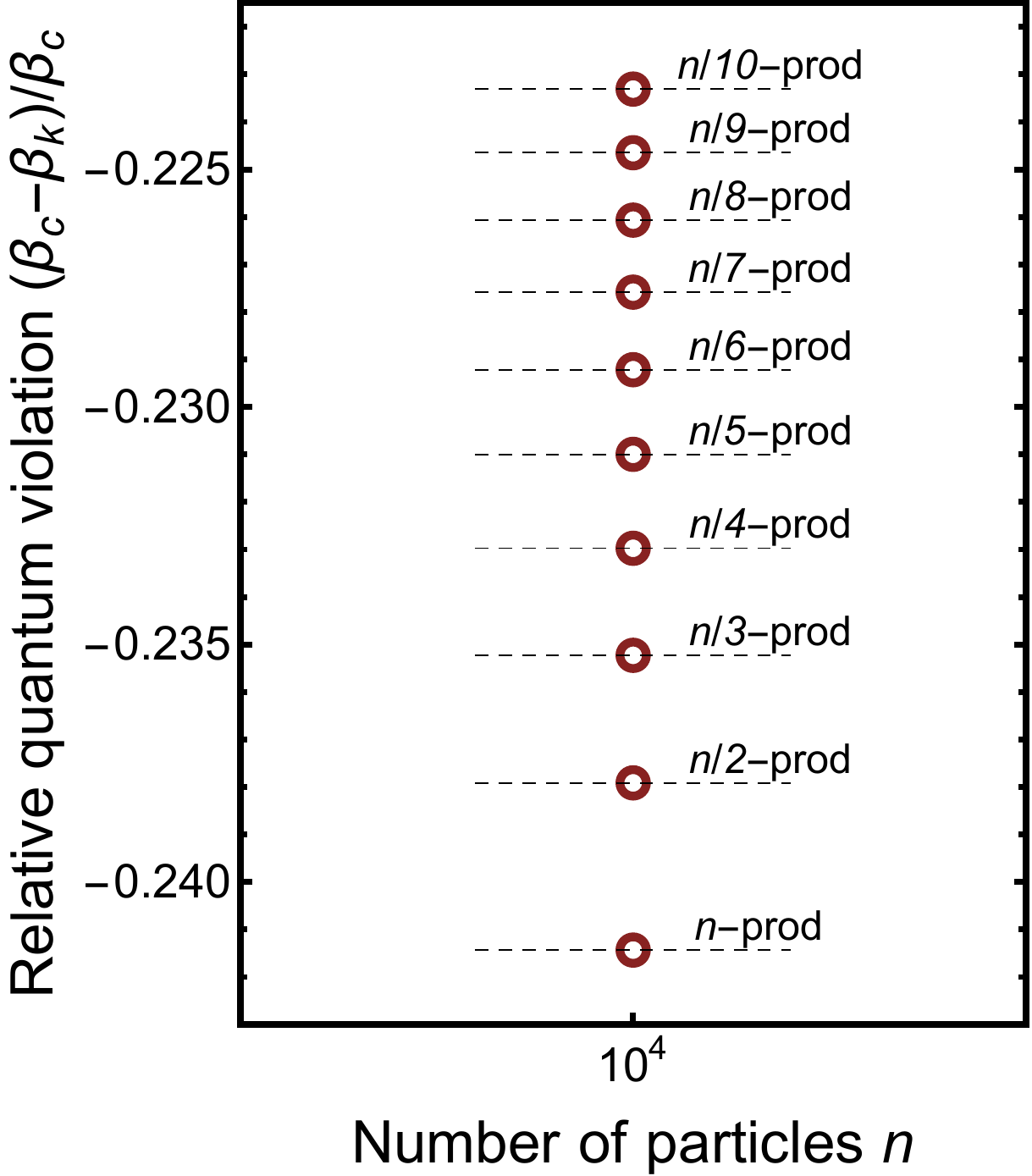}
  \caption{Asymptotic approximation of the DIWED bounds for the 2-body PIBI \eqref{eq:ineq6} with $n=10^4$. Each point corresponds to the $k$-producible bound with $k=n/m$ and $m\in\{1,\ldots,10\}$. The dotted lines are for illustrative purposes. In their derivation, we used that for sufficiently large $n$ and $k$ the optimal $k$-producible state is well approximated by a product of Gaussian superpositions of Dicke states \eqref{eq:gaussianDicke} (see \cite{PRATwin} for details).}
  \label{fig:levels}
  \end{center}
\end{figure}

\paragraph{Asymptotic behavior.}
After suitable local unitary transformations, the optimal $k$-producible state for the expression (\ref{eq:ineq6}) and sufficiently large $k$, can be well approximated analytically by a product (w.r.t. a partition $\cal P$) of Gaussian superpositions of Dicke states, each with different parameters $\mu_{\cal A}$, $\sigma_{\cal A}$:
\begin{equation}\label{eq:gaussianDicke}
 \ket{\Psi} = \bigotimes_{\cal A \in \cal P}\left(\sum_{0 \leq k_{\cal A} \leq |\cal A|} \psi_{k_{\cal A}}^{\cal A} \ket{D_{|\cal A|}^{k_{\cal A}}}\right),
\end{equation}
where $ \psi_{k_{\cal A}}^{\cal A} := e^{-(k_{\cal A}-\mu_{\cal A})^2/4 \sigma_{\cal A}}/\sqrt[4]{2\pi \sigma_{\cal A}}$. Note that, when ${\cal P} = \{[n]\}$, one recovers the analytical form of the state maximally violating \eqref{eq:ineq6} \cite{AnnPhys}. This enables us to obtain an asymptotic form for the $k$-producible bounds. For large $n$, one can well approximate $\braket{\Psi|{\mathcal{B}}|\Psi}$ by a quartic polynomial in $\mu_{\cal A}$, $\sigma_{\cal A}$ (see \cite{PRATwin} for details). In Fig. \ref{fig:levels} we show how one can gain information about the entanglement depth of the system by simply looking at the Bell inequality violation.

\begin{figure}
{{\includegraphics[width=0.50\linewidth]{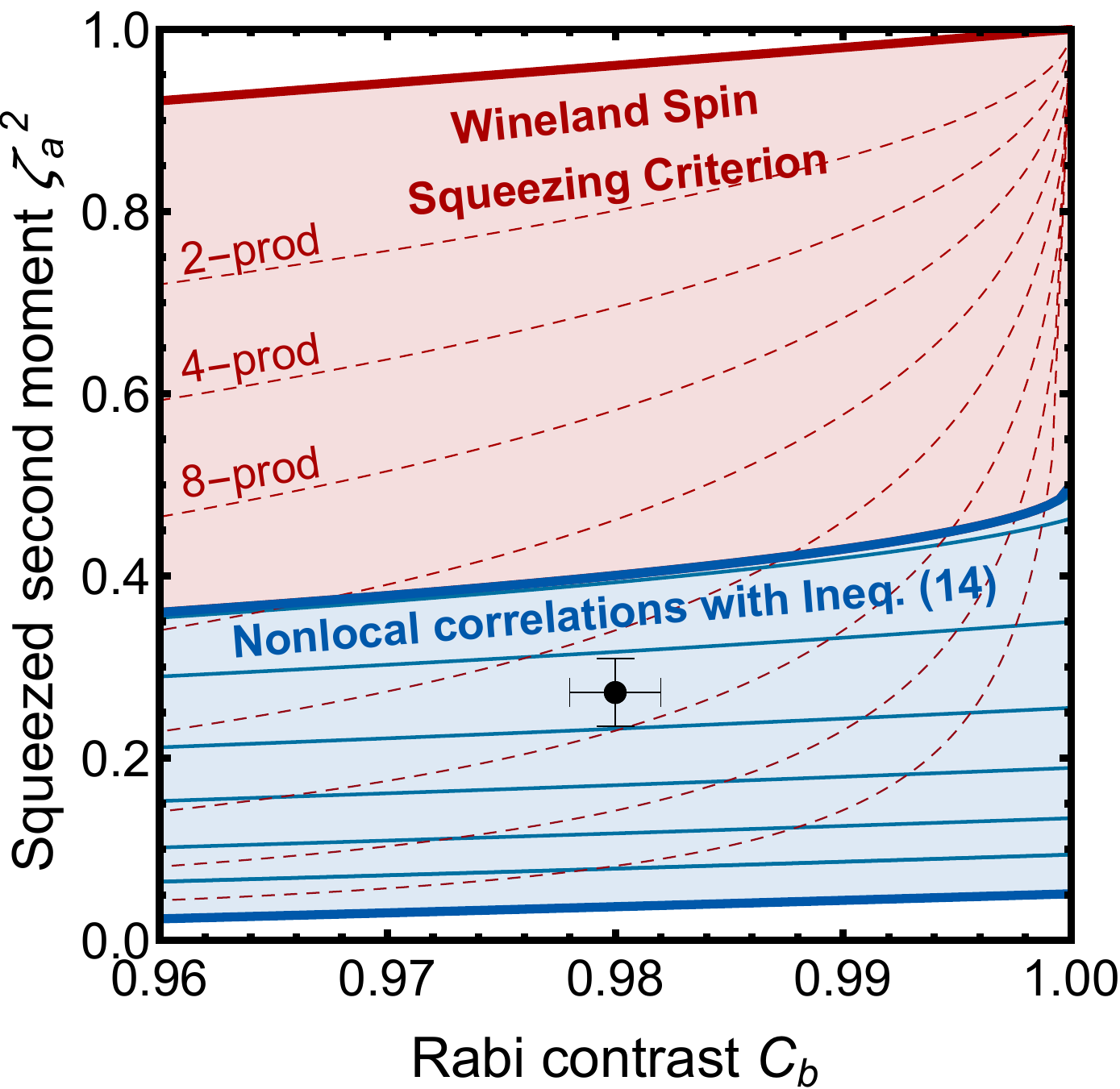}}\label{fig:collective1}}%
{{\includegraphics[width=0.50\linewidth]{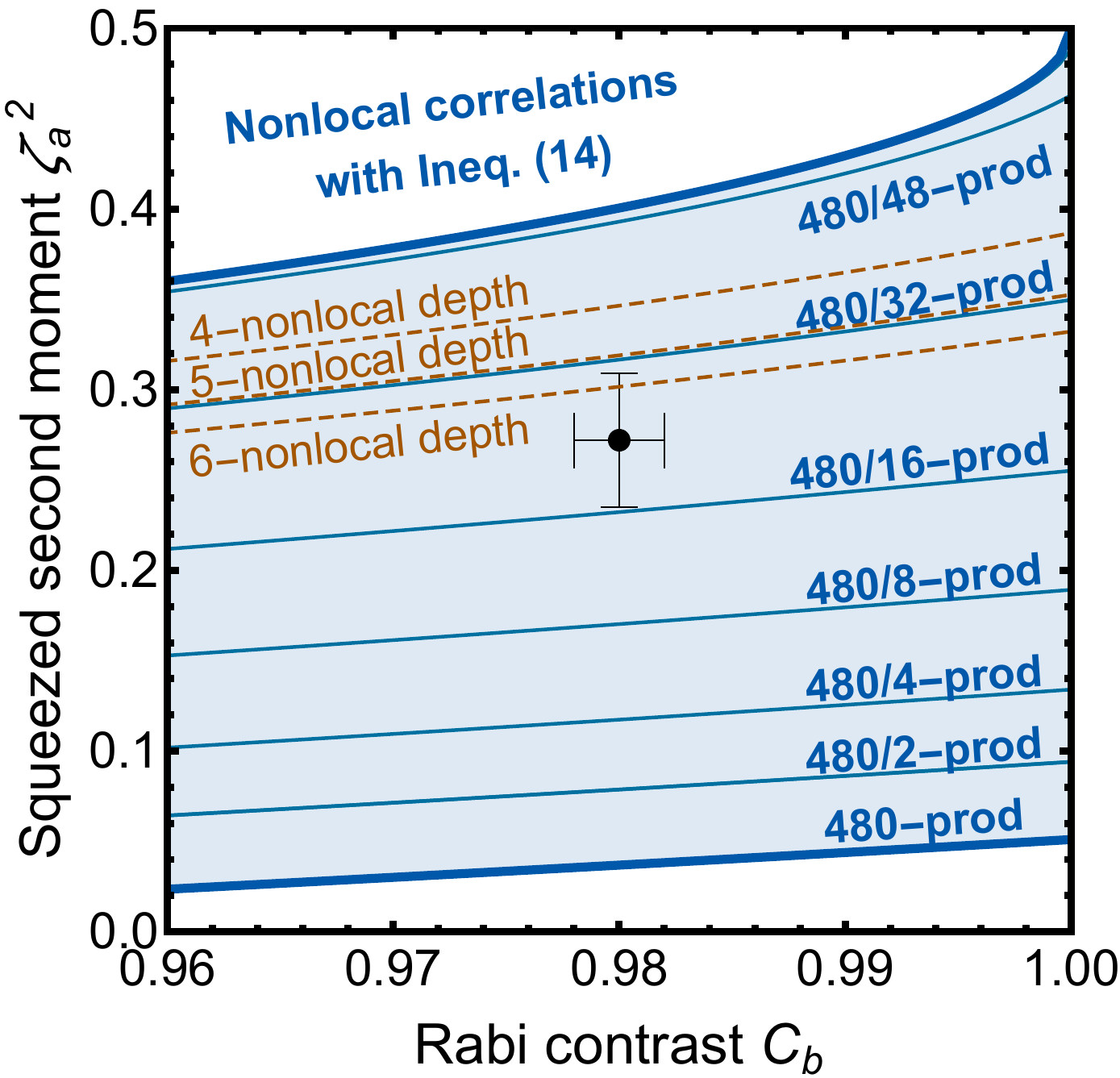}}\label{fig:collective2}}%
  \caption{Witnesses of entanglement depth from collective measurements. In order to compare our results with other known criteria and experimental data we consider $n=480$ and express the witnesses in terms of the Rabi contrast $C_b$ and the squeezed second moment $\zeta^2_a$. The area below a curve denotes violation of the corresponding witness. The black dot corresponds to the experimental data reported in \cite{ExperimentBasel} with $1$ standard deviation error bars. The left plot shows the witnesses of entanglement depth resulting from the Wineland spin squeezing criterion \cite{WinelandPRA94,SorensenMolmer2001} and the Bell correlation witness \eqref{eq:ineq6} from \cite{ExperimentBasel}. On the right, we show different $k$-producible bounds for the DIWED stemming from \eqref{eq:ineq6} (see \cite{PRATwin} for the detailed derivation) and in yellow we show the nonlocality depth witnesses derived in \cite{FlavioNonlocalityDepth} also from \eqref{eq:ineq6}. The DIWEDs presented in this work certify an entanglement depth of $15$, in comparison with the Bell correlations depth of $5$ certified in \cite{FlavioNonlocalityDepth} and the entanglement depth of $28$ certified with spin squeezing \cite{SorensenMolmer2001,ExperimentBasel}.}
  \label{fig:comparison}
\end{figure}

\paragraph{Comparison to other entanglement depth criteria and experimental data.}
One of the key features of the PIBIs from \cite{SciencePaper, AnnPhys, ThetaBodies} is that they can be effectively evaluated via a Bell correlation witness that only requires estimation of first and second moments of the total spin components. This has already been performed experimentally in 480 $^{87}$Rb atoms \cite{ExperimentBasel} and in a thermal ensemble of $5 \cdot 10^5$ atoms \cite{ExperimentKasevich}. Witnesses of entanglement depth (although not DI) based on spin-squeezing inequalities have allowed to detect $k\geq 28$ in a $8\cdot 10^3$ atom BEC \cite{LueckePRL2014}. Fig. \ref{fig:comparison} compares our DIWEDs with other entanglement depth criteria, such as the Wineland spin squeezing criterion \cite{WinelandPRA94, SorensenMolmer2001} and Bell correlation depth witnesses \cite{FlavioNonlocalityDepth}. Finally, we see that with the experimental data from \cite{ExperimentBasel}, our DIWED guarantees entanglement depth of $k \geq 15$.

\paragraph{Conclusions.}
In this letter, we have presented a method to construct DIWEDs for many-body Bell inequalities. When these DIWEDs are built from two-body PIBIs we can numerically find their $k$-producible bounds. We have tested our method against real experimental data and we see, not surprisingly, that the entanglement depth detected by our DIWEDs is larger than Bell correlation depth winesses against no-signalling resources, which are much more demanding \cite{FlavioNonlocalityDepth}, yet smaller when compared to non-DI witnesses of entanglement depth \cite{LueckePRL2014, VitaglianoNJP2017}. Interestingly, the DIWEDs proposed here can be tested within current technology, solving an open question posed in \cite{YCLiangEntDepth}, thus making them experimentally more appealling than existing DI entanglement witnesses \cite{BancalDIGME, YCLiangEntDepth}. Our method goes beyond those solely based on the NPA hierarchy, which is impractical for a larger number of parties, and those focused in GME detection, which may be too demanding technologically.

\paragraph{Acknowledgments.}
This project has received funding from the European Union's Horizon 2020 research and innovation programme under the Marie-Sk{\l}odowska-Curie grant agreement No 748549, Spanish MINECO (FISICATEAMO FIS2016-79508-P, QIBEQI FIS2016-80773-P, Severo Ochoa SEV-2015-0522, Severo Ochoa PhD fellowship), Fundacio Cellex, Generalitat de Catalunya (SGR 1341, SGR 1381 and CERCA/Program), ERC AdG OSYRIS and CoG QITBOX, EU FETPRO QUIC, the AXA Chair in Quantum Information Science.
J. T. acknowledges support from the Alexander von Humboldt foundation. R.~A.~acknowledges the support from the Foundation for Polish Science through the First Team project (First TEAM/2017-4/31) co-financed by the European Union under the European Regional Development Fund.

\vfill

\bibliography{bibliography}

\end{document}